\documentstyle[12pt]{article}
\advance \topmargin by -\headheight
\advance \topmargin by -\headsep
     
\evensidemargin \oddsidemargin
\def\rf#1{(\ref{eq:#1})}
\def\lab#1{\label{eq:#1}}
\def\nonu{\nonumber}
\def\br{\begin{eqnarray}}
\def\er{\end{eqnarray}}
\def\be{\begin{equation}}
\def\ee{\end{equation}}

\def\({\left(}
\def\){\right)}
\def\v{\vert}

\def\sskp{\par\vskip 0.15cm \par\noindent}
\def\bc{\begin{center}}
\def\ec{\end{center}}

\def\lbf#1{{\large {\bf {#1}}}}
\def\tr{\mathop{\rm tr}}

\def\a{\alpha}
\def\b{\beta}
\def\d{\delta}

\def\g{\gamma}

\def\h{{1\over 2}}
\def\l{\lambda}

\def\o{\over}

\def\pa{\partial}

\def\t{\tau}

\def\ba{\be\begin{array}{c}}
\def\ea{\end{array}\ee}

\def\cD{{\cal D}}

\def\cJ{{\cal J}}

\def\cL{{\cal L}}

\font \msb=msbm10 scaled \magstep1

\newcommand{\IC}{\mbox{\msb C} }

\def\one{\hbox{{1}\kern-.25em\hbox{l}}}
\def\0#1{\relax\ifmmode\mathaccent"7017{#1}%
        \else\accent23#1\relax\fi}

\newfam\euffam
\font\sixeuf=eufm6
\font\eighteuf=eufm8
\font\twelveeuf=eufm10 scaled\magstep1
	\textfont\euffam=\twelveeuf
	\scriptfont\euffam=\eighteuf
  	\scriptscriptfont\euffam=\sixeuf

\def\proof{\par{\it Proof}. \ignorespaces} \def\endproof{{$\Box$}\par}

\newcommand{\nit}{\noindent}
\newcommand{\ct}[1]{\cite{#1}}

\begin{document}

\nit
{\Large {\bf Integrable Structure behind WDVV Equations}}
\begin{center}
\begin{minipage}[t]{70mm}
{\bf Henrik Aratyn}\\
\\
Department of Physics,\\
University of Illinois at Chicago,\\
845 W. Taylor St.,\\
Chicago, IL 60607-7059\\
e-mail: aratyn@uic.edu\\
\end{minipage}
\begin{minipage}[t]{70mm}
{\bf Johan van de Leur}\\
\\
Mathematical Institute,\\
University of Utrecht,\\
P.O. Box 80010, 3508 TA Utrecht,\\
The Netherlands\\
e-mail: vdleur@math.uu.nl
\end{minipage}
\end{center}

\begin{center}
{\bf Abstract}
\end{center}
An integrable structure behind 
Witten--Dijkgraaf--Verlinde--Verlinde (WDVV) equations is identified with
reduction of a Riemann-Hilbert problem for a homogeneous 
$\widehat{GL} (N, \IC)$ loop group.
Reduction requires the dressing matrices to be fixed points of a loop group automorphism of 
order two resulting in a  sub-hierarchy of $\widehat{gl} (N, \IC)$ hierarchy
containing only odd symmetry flows. 
The model possesses Virasoro symmetry and imposing Virasoro constraints
ensures homogeneity property of the
Darboux-Egoroff structure.
Dressing matrices of the reduced model provide solutions of
the WDVV equations.

\sskp
\lbf{1.~Introduction}\sskp
Massive topological field theories can be  classified locally by the
Darboux--Egoroff
system of differential equations:
\begin{equation}
    \frac{\partial }{ \partial u_k} \beta_{ij} = \beta_{ik} \beta_{kj},
\;\; i \ne k \ne j , \;\;\;
\sum_{k=1}^{N} \frac{\partial }{ \partial u_k} \beta_{ij}     =0,
\;\; i \ne j \, ,
\lab{betas-comp}
\end{equation}
\begin{equation}
\beta_{ij}=\beta_{ji}\, ,
\lab{betas-sym}
\end{equation}
where in addition one also assumes that
\begin{equation}
\sum_{k=1}^Nu_k\frac{\partial }{ \partial u_k} \beta_{ij} = -\beta_{ij}\, .
\lab{betas-deg}
\end{equation}
This was observed by B. Dubrovin 
in \cite{Du3a}, \cite{Du1}.
The $\b_{ij}$, $1\le i,j\le N$ with $i\ne j$ are 
called the ``rotation coefficients''.  To be more precise, these Darboux--Egoroff 
equations are the compatibility conditions of the following linear 
system of differential equations depending on a spectral parameter $\lambda$:
\begin{eqnarray}
\frac{\partial }{ \partial u_i} A_{j\, k} ({ u}, \lambda)  &=& 
\beta_{ji} ({ u}) \,
A_{i\, k}  ({ u}, \lambda) 
\;\; ;\;\; i \ne j=1,{\ldots},N \lab{sigj}\\ 
\sum_{j=1}^{N} \frac{\partial }{ \partial u_j}  
A_{i\, k} ({ u}, \lambda)  &=& \lambda \,
A_{i\, k}  ({ u}, \lambda)  \;\; ;\;\; i =1,{\ldots},N \lab{sumsigj}
\end{eqnarray}
for each $k=1,{\ldots} , N$. 
Let $V(u)=(\beta_{ij}(u)(u_j-u_i))_{1\le i,j\le n}$ and assuming that 
this matrix is diagonalizable, one can construct a local semisimple 
Frobenius manifold, which provides an algebraic formulation of a 
massive topological field theory as follows. 
Find an invertible power series 
\begin{equation}
A(u,\lambda)= A_0(u)+ A_1(u)\lambda+A_2(u)\lambda^2+\cdots\, ,
\end{equation}
which is a solution of this system, then 
\begin{equation}
\sum_{k=1}^N u_k\frac{\partial}{\partial u_k}A_0(u)=V(u)A_0(u).
\end{equation}
It is then straightforward to check that $A_0(u)^{-1}V(u)A_0(u)$ is a constant matrix, 
hence there exists an invertible complex matrix $S$ such that 
\be
\mu=\sum_{i=1}^N \mu_i E_{ii}=S^{-1}A_0(u)^{-1}V(u)A_0(u)S
\ee
is a diagonal matrix. Now let $A(u)=A_0(u)S$ with
$A(u)=(a_{ij}(u))_{1\le i,j\le N}$, then 
\be 
\sum_{k=1}^N u_k\frac{\partial}{\partial u_k}A(u)=A(u)\mu.
\ee
Dubrovin then showed \cite{Du1} that there exists on the domain
$u_i \ne u_j$ and
$ a_{11} a_{21} \cdots a_{N1}\ne 0$ a local semisimple Frobenius 
manifold with scaling dimensions $\mu_\alpha-\mu_1$ (see also \ct{Dubrovin:1998fe}).
The diagonal metric is given by
\begin{equation}
d\, s^2=\sum_{i=1}^n a_{i1}^2(u)(du_i)^2.
\end{equation}
The flat coordinates $x^\alpha$ are determined by
\begin{equation}
\sum_\beta \eta_{\alpha\beta}\frac{\partial x^\beta(u)}{\partial u_i}
=a_{i1}(u)a_{i\alpha}(u),
\end{equation}
where
\begin{equation}
\eta_{\alpha\beta}=\sum_{i=1}^n
\a_{i\alpha}(u)\a_{i\beta}(u).
\end{equation}
Finally, the structure constants 
\begin{equation}
c_{\alpha\beta\gamma}(x(u))=
\frac{\partial^3 F(x(u))}{\partial x^\alpha \partial x^\beta \partial x^\gamma}=
\sum_{i=1}^n
\frac{a_{i\alpha}(u)a_{i\beta}(u)a_{i\gamma}(u)}{a_{i1}(u)}
\end{equation}
are the third order derivatives of the WDVV prepotential $F(x)$.

In this paper, we use a Riemann--Hilbert problem for the loop group of $GL(N)$ 
to obtain a similar structure. The symmetry conditions \rf{betas-sym}
for the rotation coefficients 
are obtained by using the Cartan involution on the loop group. 
The model possesses Virasoro symmetry and imposing Virasoro constraints
ensures the homogeneity property \rf{betas-deg} of the
Darboux-Egoroff structure.  

\newpage
\lbf{2.~Symmetry flows, Riemann-Hilbert problem and the ${\bf \tau}$
function}
\sskp

We will introduce the $G= \widehat{GL} (N)$
symmetry flows  in terms of (extended) Riemann-Hilbert 
problem involving two subgroups of 
the Lie loop group $G$ defined as :
\be
G_{-} = \left\{ g \in G \v g(\l) = 1 + \sum_{i<0} g^{(i)}  \right\}
,\;\; G_{+} = \left\{ g \in G \v g(\l) = \sum_{i\geq0} g^{(i)}  \right\}\, , 
\lab{g-pos}
\ee
here $g_i$ has grading $i$ with respect to a homogeneous gradation
defined by a derivation $\l d /d \l$.
It also holds that $G_{+} \cap G_{-} = I$.
Let the loop algebra corresponding  to $G$ be ${\widehat {\cal G}} = 
{\widehat {gl}} (N)$.
This algebra splits into the direct sum ${\widehat {\cal G}}={\widehat {\cal G}}_{+} 
\oplus{\widehat {\cal G}}_{-}$
with respect to the homogeneous gradation, where ${\widehat {\cal G}}_{\pm}$ are Lie
algebras associated with the subgroups $G_{\pm}$.

Let the ``un-dressed'' wave (matrix) function be
\begin{equation}
\Psi_0 ({\bf u}, \lambda) = \exp\left({\sum_{j=1}^{N} \sum_{n=1}^{\infty} 
E^{(n)}_{jj}u^{(n)}_j}\right)
=\sum_{j=1}^{N} E_{jj} e^{\sum_{n=1}^{\infty} \lambda^{n} u^{(n)}_j}
\lab{dfirst}
\end{equation}
We now define an extended Riemann-Hilbert factorization problem for the
homogeneous gradation :
\begin{equation}
\Psi_0 ({\bf u}, \lambda) g  (\lambda) \,
= \Theta^{-1} ({\bf u},\l) \,  M  ({\bf u},\l) 
\lab{rh-def}
\end{equation}
where $g:S^1 \to G_{-}G_{+}$
while $ \Theta \in G_{-}$, $M \in G_{+}$.
We use the multi-time notation with
$({\bf u})= ({\bf u_1}, {\ldots} , {\bf u_{N}})$
to denote $N$ multi-flows ${\bf u_j}$. Each argument
${\bf u_j}$, $j=1,{\ldots} ,N$  is an abbreviated notation for the 
multi-flows $u_j^{(n)}$ with $n$ running from $1$ to $\infty$.

For the symmetry flows one derives 
from \rf{rh-def} expressions:
\br
\frac{\partial}{\partial u^{(n)}_j} \Theta ({\bf u}, \lambda) 
&=& - \left(\Theta E^{(n)}_{jj} \Theta^{-1} 
\right)_{-} \Theta ({\bf u}, \lambda) 
\lab{uthpos}\\
\frac{\partial}{\partial u^{(n)}_j} M ({\bf u}, \lambda) 
&=&  \left(\Theta E^{(n)}_{jj} \Theta^{-1} 
\right)_{+} M ({\bf u}, \lambda) 
\lab{ummpos}
\er
where $({\ldots} )_{\pm}$ denote the projections onto ${\widehat {\cal G}}_{\pm}$.
Note, that $\sum_{j=1}^N {\partial}\Theta / {\partial u^{(n)}_j} =0$
for any $n>0$.

In the following we will denote for brevity $u_j = u^{(1)}_j$
and $ \partial_j = {\partial}/ {\partial u^{(1)}_j}$ for
$j=1,{\ldots}, N$.

It is  natural to associate to the Riemann-Hilbert factorization 
a one-form :
\be
\cJ = - \sum_{j=1}^{N} \sum_{n=1}^{\infty} 
{\rm Res}_{\l} \(\tr \(  {\Theta}^{-1}  
{d  \Theta \o d \l} E^{(n)}_{jj}  \)  \) d u^{(n)}_j
= - {\rm Res}_{\l} \(\tr \(  {\Theta}^{-1}  
{d  \Theta \o d \l} \Psi_0^{-1} d \Psi_0  \)  \) 
\lab{jnj-tau}
\ee 
which due to equation \rf{uthpos} is closed :
$ d \cJ = 0$ with $d= \sum_{j=1}^{N} \sum_{n=1}^{\infty} 
d u^{(n)}_j \frac{\pa}{\pa u^{(n)}_j}$.
Hence, locally, $ \cJ$
can be written as $d$ of a zero-form 
being some function of the dynamical variables of the model.
Conventionally, this function is denoted as the logarithm of the 
$\tau$-function:
\be
\cJ = - d  \log \tau = - \sum_{j=1}^{N} \sum_{n=1}^{\infty}
\frac{\pa  \log \tau }{\pa u^{(n)}_j}\; d u^{(n)}_j 
\lab{jnjta}
\ee
This equation can effectively be integrated to yield an explicit expression 
for the $\tau$-function. To do this it is required to extend the formalism
from the loop groups to the centrally extended groups.
As in case of the AKNS hierarchy with the $\tau$-function 
defined in \ct{Wilson:1993} 
we embed the framework in the centrally extended
group with elements denoted as ${\widehat g}$ and propose :
\be
\tau \( {\bf u}\)= \langle 0 |   
\widehat{ \Psi_0  g}  \, | 0 \rangle 
\lab{tau-def}
\ee
where $| 0 \rangle$ is the highest weight state such that $X_{\geq 0} | 0 \rangle =0$
and $\langle 0 | X_{\leq 0}=0$ and furthermore 
$\langle 0 | {\hat k} | 0 \rangle=1$, where ${\hat k}$ denotes
the direction associated to the central element within the affine
Kac-Moody algebra.
 
Let $\theta^{(-1)} $ be the grade $-1$ term of expansion:
\be
\Theta = 1 + \theta^{(-1)} \l^{-1} + \theta^{(-2)} \l^{-2}+{\ldots} \,\,.
\lab{grad-exp}
\ee
{}From equations \rf{jnj-tau} and \rf{jnjta} (for $n=1$)  one finds :
\be
(\theta^{(-1)})_{ii} = - \pa_i  \log \t \,.
\lab{thm1t}
\ee
Introduce now the ``rotation cefficients'' $\b_{ij}$ with $i\ne j$ as
the off-diagonal elements of $\theta^{(-1)}$ :
\be
\b_{ij}\, =\, (\theta^{(-1)})_{ij} , \;\;\;\; i\ne j=1,{\ldots} ,N
\lab{them1}
\ee
For $ \theta^{(-1)}$ matrix we find from \rf{uthpos} :
\be
\pa_{j}  \theta^{(-1)}  = 
\lbrack \,   E_{jj}   \, , \, \theta^{(-2)}  
\, \rbrack \  + \lbrack \, \theta^{(-1)} \, , \,  E_{jj} 
\, \rbrack \, \theta^{(-1)}
\lab{pajthm1}
\ee
from which it follows that :
\be
\pa_{j}  (\theta^{(-1)})_{ik} = \( \theta^{(-1)}\)_{ij} 
(\theta^{(-1)})_{jk}, 
\quad  \quad j\ne i,\, \,j \ne k\,.
\lab{pajthik}
\ee
Accordingly, the ``rotation coefficients'' $\beta_{ij}$ given by \rf{them1}   
satisfy \rf{betas-comp}.

As found in \ct{Aratyn:2001cj}, the matrix elements 
of the dressing matrix  $\Theta=(\theta_{ij})_{1\le i,j\le N}$ are
given by:
\be
\theta_{i\,j} ({\bf u}, \lambda) = \frac{1}{\tau ({\bf u})} \times  \left\{
\begin{array}{lr}
 \frac{1}{\lambda} \beta_{i\,j} ({\bf u_j}-[\lambda^{-1}]) {\tau
({\bf u_j}-[\lambda^{-1}])}  &i \ne j \\
  {\tau ({\bf u_i}-[\lambda^{-1}])} &
i =j \end{array}\right.
\lab{th-gen}
\ee
where the multi-flow arguments ${\bf u_j}$ were shifted
according to $({\bf u_j}-[\lambda^{-1}])= (u_j^{(1)}-1/\lambda,
u_j^{(2)}-1/2\lambda^2,{\ldots} )$.

The matrix coefficients of the wave matrix $\Psi$:
\begin{equation}
\Psi ({\bf u}, \lambda) = \Theta ({\bf u}, \lambda)\, \Psi_0 ({\bf u}, \lambda)  
\lab{gms}
\end{equation}
and the matrix coefficients of $M({\bf u},\lambda)=\Psi ({\bf u}, \lambda)
g(\l)$ satisfy the linear system \rf{sigj}-\rf{sumsigj},    
which are the compatibility conditions of 
equations \rf{betas-comp}. This is the
Darboux-Egoroff system provided we also impose the symmetry condition
\rf{betas-sym} on the rotation coefficients.
To do this we reduce the integrable hierarchy associated with the flow
equation equations \rf{uthpos},\rf{ummpos}
using the automorphism  $\sigma$ :
\be
\sigma \( X ( \l ) \)  = \( \( X( -\l ) \)^T \)^{-1}, \; \quad
X \in G= \widehat{GL} (N)
\lab{lgsauto}
\ee
which leaves the evolution equations \rf{uthpos}-\rf{ummpos}
invariant for odd flows only (labeled by $n$ being an odd integer).
Accordingly, we define the integrable sub-hierarchy by constraining the
dressing matrices $\Theta ({\bf u}, \l )$ and
$ M ({\bf u}, \l )$  to the fixed points of the
loop group  automorphism $\sigma $ \rf{lgsauto} :
\be
\Theta^{-1} ( {\bf u}, \lambda ) =     \Theta^T ( {\bf u}, -\lambda ), \quad
M^{-1} ( {\bf u}, \lambda ) =    M^T ( {\bf u}, -\lambda )
\lab{red-thet}
\ee
with $\Theta ({\bf u}, \l )$ and
$ M ({\bf u}, \l )$ depending only on odd coordinates ${\bf u}$:
( $u^{(2k+1)}_j, k=0,1, {\ldots} $).
The odd flows of the reduced 
sub-hierarchy preserve the conditions \rf{red-thet}.
The fixed points of the automorphism $\sigma$ form a subgroup of
$G= \widehat{GL} (N)$, called a twisted loop group of $GL (N)$
(see e.g. \ct{vandeLeur:2000gk}).
For the first term $\theta^{(-1)}$ of the expansion \rf{grad-exp}
the constraint \rf{red-thet} implies that 
$\theta^{(-1)} = \theta^{(-1)\, T}$.
Accordingly, the ``rotation coefficients'' $\beta_{ij}$ associated to the
subhierarchy via \rf{them1} are symmetric and give rise to the 
Darboux-Egoroff metrics.

\sskp
\lbf{3.~Condition of conformal invariance and homogeneity}
\sskp
We propose here the Virasoro constraints in the setting of the
Riemann-Hilbert problem approach to the integrable
models.
In this setting the
Virasoro constraints are seen to arise as symmetries of the
hierarchy. 

The action of the Virasoro symmetry flows on the dressing
matrices is given by:
\br 
\delta^V_k  \, \Theta ({\bf u}, \lambda) 
&=&  (\Theta L_k \Theta^{-1})_{-} \Theta 
=(\Theta l_k \Theta^{-1})_{-} \Theta 
- \sum_{j=1}^{N} \sum_{n=1}^{\infty} n  u_j^{(n)}
\frac{\pa \Theta}{  \pa u_j^{(n+k)}} 
\lab{th-virasoro} \\
\delta^V_k   M ({\bf u}, \lambda) 
&=& - (\Theta L_k \Theta^{-1})_{+} M 
=- (\Theta l_k \Theta^{-1})_{+} M 
- \sum_{j=1}^{N} \sum_{n=1}^{\infty} n  u_j^{(n)}
\frac{\pa  M}{ \pa u_j^{(n+k)}} 
\lab{m-virasoro}
\er
where
\br
L_k& =&  - \lambda^{k+1} {d \o d \lambda} +  \sum_{j=1}^{N} \sum_{n=1}^{\infty} n  u_j^{(n)}
E_{jj}^{(k+n)}, \quad k \geq -1
\lab{lkvir} \\
& =&  - \lambda^{k+1} \frac{d}{d \lambda} +  
 \sum_{n=1}^{\infty} n  U^{(n)} \l^{k+n} , \nonu\\
 U^{(n)} & =& {\rm diag} \( u^{(n)}_1, {\ldots} ,u^{(n)}_N\)=\sum_{i=1}^N
 u^{(n)}_i E_{ii} \nonu
\er
and $l_k = - \lambda^{k+1} d / d \lambda$.
Both $l_k$ and $L_k$ for $k\ge -1$  generate a subalgebra of the Witt (centerless Virasoro) 
algebra:
\be
\lbrack {l_k}, {l_r}\rbrack = (k-r) l_{k+r}, \quad
\lbrack {L_k}, {L_r}\rbrack = (k-r) L_{k+r}
\lab{vir-xmxn}
\ee
{}The signs in eqs. \rf{th-virasoro} and \rf{m-virasoro} have been chosen
in such a way that the transformation $\delta^V_k $ also satisfies the Virasoro algebra
$ \lbrack \delta^V_k , \delta^V_r\rbrack = (k-r) \delta^V_{k+r} $
when applied on the dressing matrices $\Theta$ and $M$.
Furthermore, the transformations $\delta^V_k $ commute with the flows 
$\pa / \pa  u_j^{(n)}$:
\be
\lbrack \delta^V_k , \frac{\pa}{\pa  u_j^{(n)}} \rbrack
\Theta ({\bf u}, \lambda) =0
\lab{comdvu}
\ee
Accordingly, the Virasoro transformations  from eqs. \rf{th-virasoro} and 
\rf{m-virasoro} constitute additional symmetries of the integrable model 
defined by the Riemann-Hilbert factorization problem.
One can show that only the even Virasoro flows 
will preserve the conditions \rf{red-thet}
and define additional symmetries of the reduced 
sub-hierarchy associated with the twisted loop group of $GL (N)$.
We can therefore consistently  impose the constraints
$\d^V_k  \, \Theta ({\bf u}, \lambda) =0$ with even $k$.
For $k=0$ the equation \rf{th-virasoro} specializes to:
\be 
\d^V_0  \, \Theta ({\bf u}, \lambda) 
= \l \frac{d}{d \l} \Theta 
- \sum_{j=1}^{N} \sum_{n=1}^{\infty} n  u_j^{(n)}
{\pa \Theta  \o \pa u_j^{(n)}} 
\lab{z-virasoro} 
\ee
as follows from identity $-(\Theta \l d \Theta^{-1}/{d \l})_{-} \Theta 
= \l {d}\Theta /{d \l} $.
Hence for $k=0$ the constraint $\d^V_k  \, \Theta ({\bf u}, \lambda) =0$
takes a form 
\be
\( \l \frac{d}{d \l} - E_0 \)  \Theta ({\bf u}, \lambda) = 0
\lab{confrob}
\ee
where we have introduced the Euler vector field $E_0$:
\be
E_0 = \sum_{j=1}^{N} \sum_{n=1}^{\infty} n  u_j^{(n)}
 {\pa    \over \pa u^{(n)}_j}
\lab{euler}
\ee
The homogeneity condition \rf{confrob} is also a condition for 
the conformal Frobenius manifold.

By projecting in equation \rf{z-virasoro} on  the terms with $-1$ grade 
we obtain :
\be
\delta^V_0  \, \theta^{(-1)} 
= - \theta^{(-1)} - \sum_{j=1}^{N} \sum_{n=1}^{\infty} n  u_j^{(n)}
{\pa \theta^{(-1)} \o \pa u_j^{(n)}}
\lab{dvkztm1}
\ee
where $\theta^{(-1)} $ is the grade $-1$ term of expansion:
\rf{grad-exp}.

Hence, the  condition \rf{confrob} implies that 
\be
E_0 \,\, \theta^{(-1)}\,= \,- \theta^{(-1)}
\lab{vnottheta}
\ee
Plugging relations \rf{thm1t} and \rf{them1} into 
equation \rf{dvkztm1} we obtain :
\br
\delta^V_0 \pa_i \log \tau  &=&   \pa_i  \delta^V_0 \log \tau =
- \pa_i \sum_{j=1}^{N} \sum_{n=1}^{\infty} n  u_j^{(n)}
 {\pa  \log \tau  \over \pa u^{(n)}_j} , \quad i=1, {\ldots} ,N
\lab{delvzz}\\
\delta^V_0 \b_{kl} &=&  -\b_{kl} - \sum_{j=1}^{N} \sum_{n=1}^{\infty} n  u_j^{(n)}
 {\pa  \b_{kl}  \over \pa u^{(n)}_j}
\lab{delvb}
\er

\sskp
\lbf{4.~Virasoro flows on the ${\bf \tau}$ function}
\sskp
This subsection studies action of the general Virasoro transformation
on the $\tau$-function of the underlying integrable system.

Recall an expression for derivatives of $\log \tau$ 
given in \rf{jnj-tau} and \rf{jnjta}.
Applying $\pa_i$ on both sides of eq. \rf{jnjta}
we obtain an identity:
\be
{\pa^2  \log \tau  \over \pa u^{(n)}_j \pa u_i}=
{\rm Res}_{\l} \(\tr  \(\Theta E^{(n)}_{jj} \Theta^{-1} E_{ii}\) \)
\;\; i,j =1,{\ldots} ,N, \;\;\; n>0
\lab{idejei}
\ee

Combining eqs. \rf{th-virasoro} and \rf{jnj-tau}-\rf{jnjta} we find
\be
\delta^V_k  {\pa  \log \tau  \over \pa u^{(n)}_j}
= {\rm Res}_{\l} \(\tr \({\Theta}  E^{(n)}_{jj} {\Theta}^{-1}  
\frac{d}{ d \l} (\Theta L_k \Theta^{-1})_{-} \)  \) 
\lab{dvktan}
\ee
Since $ \l d (\l^{-1}) /d \l= - \l^{-1}$ the last equation 
becomes
\be
\delta^V_k  \pa_i  \log \tau 
= - {\rm Res}_{\l} \(\tr \(  E_{i\,i}
\Theta l_k \Theta^{-1} \)  \)
- \sum_{j=1}^{N} \sum_{n=1}^{\infty} n  u_j^{(n)}
{\rm Res}_{\l} \(\tr \(\Theta E^{(k+n)}_{jj} \Theta^{-1} E_{i\,i}\)\)
\lab{dvktab}
\ee
Using the identity \rf{idejei} we rewrite the last equation as 
\be
\delta^V_k  \pa_i  \log \tau 
= - {\rm Res}_{\l} \(\tr \(  E_{i\,i}
\Theta l_k \Theta^{-1} \)  \)
-  \sum_{j=1}^{N} \sum_{n=1}^{\infty} n  u_j^{(n)}
\pa_i {\pa  \log \tau  \over \pa u^{(k+n)}_j}
\lab{dvktaba}
\ee

The cases $k=0,1$ were calculated and eq. \rf{dvktab}
gives :
\be
\pa_i \delta^V_k \log \tau = - \pa_i 
\sum_{j=1}^{N} \sum_{n=1}^{\infty} n  u_j^{(n)}
 {\pa  \log \tau  \over \pa u^{(n+k)}_j},\;\;\; \quad k= 0,1
\lab{delvkz}
\ee
and 
\be
\pa_i \delta^V_{-1} \log \tau = - \pa_i 
\sum_{j=1}^{N} \sum_{n=2}^{\infty} n  u_j^{(n)}
 {\pa  \log \tau  \over \pa u^{(n-1)}_j} \, .
\lab{delvkmo}
\ee
Only $k=0$ case applies to the reduced case.

{}From eqs. \rf{th-virasoro} and \rf{m-virasoro} we find :
\br
\delta^V_k  \( \Theta^{-1} M \) &=&
- \Theta^{-1}  (\Theta L_k \Theta^{-1})_{-} M - 
\Theta^{-1}  (\Theta L_k \Theta^{-1})_{-} M 
= - L_k\( \Theta^{-1}M \)
\nonu \\
&=& \l^{k+1} \frac{d}{d \l} \( \Theta^{-1}M \)
- \sum_{j=1}^{N} \sum_{n=1}^{\infty} n  u_j^{(n)}
E_{jj}^{(n+k)}\( \Theta^{-1}M \) 
\lab{dvktma}
\er

Recalling the factorization problem \rf{rh-def}
we find 
\be
\delta^V_k  \( \Theta^{-1} M \) = 
\Psi_0 ({\bf u}, \lambda)  \l^{k+1} {d g \o d \l} 
\lab{dlg}
\ee
since
\be
\l^{k+1} \frac{d}{d \l} \Psi_0 ({\bf u}, \lambda)  
= \sum_{j=1}^{N} \sum_{n=1}^{\infty} n  u_j^{(n)}
E_{jj}^{(n+k)} \Psi_0 ({\bf u}, \lambda)  
\lab{lkone}
\ee
Recall, the definition \rf{tau-def} of the $\tau$-function
$\tau_g \( {\bf u}\)= \langle 0 | \widehat{ \Psi_0  g} \, | 0 \rangle$,
rewritten here in the form which explicitly shows dependence of $\tau$ on 
the loop group element $g (\l)$.

It now appears that
\be
\delta^V_k  \tau_g =\langle 0 | \l^{k+1}  \widehat{ \Psi_0 {d g \o d \l}} \, | 0 \rangle
= \tau_{\l^{k+1} {d g / d \l} }
\lab{dkvtg}
\ee
So under infinitesimal $S^1$ diffeomorpism $ g \to g+ \l^{k+1} {d g / d \l}$
the $\tau$-function transforms as
\be
\tau_g \to \tau_{g+\l^{k+1} {d g / d \l}} = \tau_g + \d^V_k \tau_g 
\lab{taugtra}
\ee

\sskp
\lbf{5.~Virasoro symmetry and dressing}
\sskp
Let 
\be
\d^{(n)}_j = {\pa  \o \pa u_j^{(n)}}  - E^{(n)}_{jj}
, \quad\;\; \d_j = \d^{(1)}_j = \pa_j - \l E_{jj}
, \quad\;\; j = 1, {\ldots} , N
\lab{delnj}
\ee
be a set of commuting operators :
\be
\lbrack \d^{(n)}_j \, , \, \d^{(k)}_i \rbrack =0,
\lab{comma}
\ee

The above operators and the Virasoro operators $L_k$ from \rf{lkvir} commute with each 
other:
\be
 \lbrack \d^{(n)}_j \, , \, L_k \rbrack =0
 , \quad i,j=1, {\ldots} , N
\lab{comm}
\ee
Both operators \rf{delnj} and \rf{lkvir} annihilate the bare wave function
$\Psi_0$ from \rf{dfirst} :
\be
L_k \Psi_0 = 0  , \quad \d^{(n)}_j \Psi_0 = 0  , \quad j=1, {\ldots} , N\,.
\lab{anni}
\ee
Similarly the operators:
\be
\cD^{(n)}_j = \Theta \d^{(n)}_j \Theta^{-1}, \quad
\cL_k = \Theta L_k \Theta^{-1},
\lab{dresop}
\ee
satisfy the commutation relations \rf{comm} and \rf{vir-xmxn}
and annihilate the wave (matrix) function $\Psi  ({\bf u}, \lambda)$ defined
in \rf{gms} :
\be
\cL_k \Psi = 0  , \quad \cD^{(n)}_j \Psi = 0  , \quad j=1, {\ldots} , N\,.
\lab{annid}
\ee

Since
\be
\Theta  \pa_j \Theta^{-1} = \pa_j + \Theta ( \pa_j \Theta^{-1} )
= \pa_j + \( \Theta \l E_{jj} \Theta^{-1} \)_{-}
\lab{tpaiti}
\ee
we obtain 
\be
\cD_j = \cD^{(1)}_j = \Theta \( \pa_j - \l E_{jj}\)\Theta^{-1} = 
\pa_j - \( \Theta \l E_{jj} \Theta^{-1} \)_{+}
= \pa_j - \l E_{jj} - V_j
\lab{tpaitia}
\ee
where 
\be
V_j \equiv \lbrack \theta^{(-1)} \, , \, E_{jj}\rbrack , \;\;\;
(V_j)_{kl}  = \( \d_{lj} -  \d_{kj}\)  \b_{kl} 
\lab{defvi}
\ee
for $\theta^{(-1)}$ defined in \rf{grad-exp}.

Consider 
\be
\l {d \Psi \o d \lambda} = \l {d \Theta \o d \lambda} \Psi_0
+ \Theta  \sum_{j=1}^{N} \sum_{n=1}^{\infty} n  u_j^{(n)}
E_{jj}^{(n)} \Psi_0
\lab{ldlp}
\ee
Recall, now from 
\rf{z-virasoro} that the condition 
$ \d^V_{k=0} \Theta ({\bf u}, \lambda) =0$  \rf{confrob} implies
\be
\l {d \Theta \o d \lambda} = 
- (\Theta  \sum_{j=1}^{N} \sum_{n=1}^{\infty} n  u_j^{(n)} E_{jj}^{(n)} 
\Theta^{-1})_{-} \Theta 
\lab{dldld}
\ee
and accordingly 
\be
\l {d \Psi \o d \lambda} = (\Theta  \sum_{j=1}^{N} \sum_{n=1}^{\infty} n  u_j^{(n)} 
E_{jj}^{(n)} \Theta^{-1})_{+} \Psi
\lab{lpsi}
\ee

Let us now study the operator
$\cL_{k=-1} = \Theta L_{-1} \Theta^{-1}$ from \rf{dresop} for $\Theta$
satisfying the homogeneity condition \rf{dldld}.
\br
\cL_{-1} &=& - \frac{d}{d \l} + {d \Theta \o d \lambda} \Theta^{-1}
+  \sum_{j=1}^{N} \sum_{n=1}^{\infty} n  u_j^{(n)} \Theta 
E_{jj}^{(n-1)} \Theta^{-1} \lab{lmnso}\\
&=& - \frac{d}{d \l} - \l^{-1}  \sum_{j=1}^{N} \sum_{n=1}^{\infty} n  
u_j^{(n)} (\Theta E_{jj}^{(n)} \Theta^{-1})_{-}
+   \sum_{j=1}^{N} \sum_{n=1}^{\infty} n  u_j^{(n)} \Theta 
E_{jj}^{(n-1)} \Theta^{-1} \nonu\\
&=& 
- \frac{d}{d \l} + \l^{-1}  \sum_{j=1}^{N} \sum_{n=1}^{\infty} n  
u_j^{(n)} (\Theta E_{jj}^{(n)} \Theta^{-1})_{+}
\nonu
\er
and for $u^{(n)}_j=0, n>1$ one obtains
\be 
\cL_{-1} =- \frac{d}{d \l} + U + \l^{-1} 
\lbrack \theta^{(-1)} \, , \, U \rbrack
= - \frac{d}{d \l} + U + \l^{-1} V
\lab{lmnsor}
\ee
The components of the matrix
$V = \lbrack \theta^{(-1)} \, , \, U \rbrack$
are
\be
V_{ij} = (u_j-u_i) \theta^{(-1)}_{ij} = (u_j-u_i) \beta_{ij}, \quad
i,j=1, {\ldots} ,N
\lab{v-comps}
\ee
It follows from the construction that
the wave (matrix) function $\Psi  ({\bf u}, \lambda)$ defined
in \rf{gms} is annihilated by $\cL_{-1} $ and $\cD_j $
\be
\cL_{-1} \Psi  ({\bf u}, \lambda) =0 \; \to \;
{d \Psi \o d \lambda} = (U + \l^{-1} V) \Psi 
\lab{lmopsi}
\ee
and 
\be
\cD_j   \Psi  ({\bf u}, \lambda) =0 \; \to \;
\frac{\pa \Psi}{\pa u_j} =(\lambda E_{jj}+V_j) \Psi \, .
\lab{cdjpsi}
\ee
Compatibility of the above equations amounts to
$\lbrack \cD_j  \, , \, \cL_{-1} \rbrack =0$,
which follows by dressing of 
$ \lbrack \d_j \, , \, L_{-1} \rbrack =0$.
{}From commutation relations $\lbrack \cD_j  \, , \, \cL_{-1} \rbrack =0$ and
$\lbrack \cD_i  \, , \, \cD_j \rbrack =0$ one obtains :
\br
\pa_j V &=& \lbrack V_j  \, , \, V\rbrack \lab{pjv} \\
\pa_j V_i &=& \pa_i V_j +\lbrack V_j  \, , \, V_i\rbrack \lab{pijv} \\
 \lbrack V  \, , \, E_{jj}\rbrack &=& \lbrack V_j  \, , \, U\rbrack 
\lab{lmodjeqs}
\er

\sskp
\lbf{6.~Isomonodromic Tau function ${\bf \tau_I}$ }
\sskp

We now consider the coefficients $ -\pa_j \log \tau $ of the one-form
$\cJ$ from \rf{jnj-tau} with the conformal constraint \rf{dldld} imposed.
Recall, that this constraint follows from the homogeneity condition 
\rf{confrob} and in view of relations \rf{delvzz} and \rf{delvb}
it holds in the reduced case that $E_0  \tau = \mu \tau $ and 
$E_0 \b_{kl} =- \b_{kl} $ where $E_0 = \sum_{j=1}^N u_j \pa / \pa u_j$
and $\mu$ is independent of variables
$u_j, j=1,{\ldots} ,N$.

Plugging relation \rf{dldld} into the defining equations \rf{jnj-tau}
and \rf{jnjta} we obtain for $n=1$:
\be
\pa_j \log \tau = -
{\rm Res}_{\l} \(\tr \(  E_{jj} 
(\Theta  \sum_{j=1}^{N} \sum_{n=1}^{\infty} n  u_j^{(n)} E_{jj}^{(n)} 
\Theta^{-1})_{-}   \)  \) , \; \, j=1, {\ldots} ,N
\lab{conf-tau}
\ee 
In the rest of the paper we set all $u_i^{(k)}=0$ for $k>1$ leaving the model
defined only in terms of the canonical coordinates 
$u_j, j=1, {\ldots} ,N$ and assume from now on that
$\sigma(g(\lambda))=g(\lambda)$ and that \rf{red-thet} holds.

The formula \rf{conf-tau} simplifies significantly and takes the form of
\be
\pa_j \log \tau = -
{\rm Res}_{\l} \(\tr \(  E_{jj} 
(\Theta \l U \Theta^{-1})_{-}   \)  \) =-
\tr \( E_{jj} \( \lbrack \theta^{(-2)} , U \rbrack + \h \lbrack \theta^{(-1)},
\lbrack \theta^{(-1)}, U  \rbrack  \rbrack  \)  \) 
\lab{conf-tbu}
\ee 
Using invariance of the trace $\tr (A \lbrack B,C\rbrack) = \tr (\lbrack A ,
B\rbrack C)$ we obtain expression:
\be
\pa_j \log \tau = \h \tr \( V_j V\)
\lab{conf-tcu}
\ee 
with $V_j, V$ matrices defined in \rf{defvi} and \rf{v-comps} and related
through the commutation relation \rf{lmodjeqs}.

The equation \rf{pjv} can be cast into  the Hamiltonian form (see also \cite{Du1})
\be
\pa_j V = \left\{V \, , \, H_j \right\}  \lab{ham-pjv}
\ee
with respect to the standard Poisson bracket $\{ \cdot \, , \, \cdot \}$
on $so (N) $ and with the Hamiltonians 
\be
 H_i=\frac{1}{2}\sum_{j\ne i}\frac{V_{ij}^2}{u_i-u_j}.
\lab{hamson} 
\ee
The commutation relation \rf{lmodjeqs} in components
gives a relation
\be
(u_k-u_j ) (V_i)_{kj} = \( \d_{ik}  - \d_{ij}\) V_{kj} , \quad k\ne j =1,
{\ldots} ,N
\lab{lmodjcom}
\ee
which can be used to prove that
\be
\h \tr \( V_i V\) = \sum_{j\ne i}\frac{V_{ij}V_{ji}}{u_i-u_j} =
- \sum_{j\ne i}\frac{V_{ij}^2}{u_i-u_j} = -2 H_i
\lab{trvjv}
\ee
{}From equations \rf{pjv} and \rf{pijv} it follows that 
$d \tr\sum_{i=1}^N  \( V_i V\) du_i =0$
and the one-form $\sum_{i=1}^N H_i du_i $ is closed.
Locally, this one-form can be represented by the derivative of the logarithm
of the so-called isomonodromy tau-function $\tau_I$ 
 \[
 d\log \tau_I=\sum_{i=1}^N H_i du_i= -\frac{1}{2}d\log \tau .
 \]
which implies that upto a multiplicative constant :
\be
\tau_I={1}/{\sqrt \tau}.
\ee
This formula was also obtained in a different way in  \cite{LM}.
\sskp
\lbf{7.~Frobenius structure}
\sskp
Expand, the dressing matrix $M ( u, \l) $ according to the grading :
\be
M ( u, \l) \, =\, M_0 ( u) + M_1 ( u) \l + M_2( u) \l^2 + {\ldots} 
\lab{m-exp}
\ee
Then, according to the condition \rf{red-thet} the zero grade term 
$M_0 (u) = M ( u, 0)= (m_{ij} (u) )_{1\le i,j \le N} $ is an orthogonal matrix
$M_0^T ( u) = M_0^{-1} ( u) $ which as seen from 
eqs. \rf{ummpos} and \rf{m-virasoro} also satisfies the flow equations :
\be
\pa_j M_0 ( u)
=\lbrack   \theta^{(-1)} ( u) \, , \, E_{jj}\rbrack M_0 ( u) = V_j ( u) M_0 ( u), \,\;\;\;
\delta^V_0 M_0 ( u) = - V ( u) M_0 ( u)
\lab{mzflows}
\ee
and hence also
\be
\sum_{j=1}^N u_j\pa_j M_0 ( u)=V ( u) M_0 ( u).
\ee
Define 
\be
{\cal V} = M_0^{-1} V M_0 \lab{defv},
\ee
it is easy to see that $\cal V$  satisfies:
\be
\pa_j {\cal V} =0 , \;\; {\cal V}^T = -{\cal V} , \ 
\delta^V_0 {\cal V} =0, \;\;\; j =1, {\ldots} , N
\lab{calvi}
\ee
and accordingly ${\cal V}$ is a constant.
We can bring ${\cal V}$ into Jordan normal form $\mu=\hat\mu +R_0$, 
with $\hat \mu$ its semisimple part and $R_0$ its nilpotent part, 
i.e. there exists a complex invertible 
matrix $S$, such that
\be
{\cal V}=S\mu S^{-1},
\ee
Now define 
\be
M(u)=M_0(u)S=(m_{ij}(u))_{1\le i,j\le N},
\ee
then using \rf{mzflows} we obtain:
\be 
\pa_j M ( u)
=\lbrack   \theta^{(-1)} ( u) \, , \, E_{jj}\rbrack M ( u) = V_j ( u) M ( u), \,\;\;\;
\delta^V_0 M ( u) = - M ( u)\mu
\lab{mzflows2}
\ee
and hence also
\be
\sum_{j=1}^N u_j\pa_j M ( u)= M ( u)\mu.
\ee
Now assume that ${\cal V}$ is diagonalizable, i.e.
the matrix  $V(u)$ has no nilpotent part, 
then
\be
\mu=\sum_{i=1}^N \mu_i E_{ii},\quad \mbox{with}\quad \mu_i=-\mu_{N+1-i}
\ee
and the columns  of $M(u)$ are eigenvectors for the Euler vectorfield $E_0$.
Hence, on the domain
$u_i \ne u_j$ and
$ m_{11} m_{21} \cdots m_{N1}\ne 0$ we have a  local semisimple Frobenius 
manifold (see e.g. \ct{Dubrovin:1998fe}) with  Lam\'e coefficients
\be 
h_i = m_{i1}. 
\ee
Define,
\be
\eta=(\eta_{\alpha\beta})_{1\le \alpha,\beta\le N}=M^TM=S^TS,
\quad \hbox{and denote}\quad \eta^{-1}=(\eta^{\alpha\beta})_{1\le \alpha,\beta\le N},
\ee
then $\mu\eta+\eta\mu=0$ and 
the derivatives with respect to the flat coordinates $x^{\a},\, \a=1,{\ldots} 
,N$ are given by
\be
\frac{\pa }{\pa x^{\a}} = 
\sum_{i=1}^N \frac{m_{i\a}}{m_{i1}} \frac{\pa }{\pa u_i} 
\lab{flatt}
\ee 
and
\be
\frac{\partial }{\partial u_i}
=\sum_{\alpha,\beta=1}^N \eta^{\alpha\beta}m_{i1}m_{i\beta}\frac{\pa }{\pa x^{\a}}.
\ee
The metric equals
\be
ds^2=\sum_{i=1}^N h_i^2(du_i)^2=\sum_{\alpha,\beta=1}^N \eta_{\a \b}dx^\alpha dx^\beta,\quad
\hbox{with}
\quad \eta_{\a \b} 
= \sum_{i=1}^N m_{i\a} m_{i\b}. 
\lab{heta}
\ee 
and the structure constants are given by
\be
c_{\a \b \g} = \sum_{i=1}^N
\frac{m_{i\a}m_{i\b}m_{i\g}}{m_{i1}}.
\lab{cstrus}
\ee
Since we have constructed everything in such a way that the columns of $M(u)$ are eigenvectors,
our Frobenius manifold has scaling dimensions $\mu_\alpha-\mu_1$.

Now, multiply the wave (matrix) function 
$\Psi  (u, \lambda)$, defined
in \rf{gms}, from the left  by 
\be
M^{-1} (u)=S^{-1}M_0^{-1}(u)=\eta^{-1}S^TM_0^{T}(u)=\eta^{-1} M^{T}(u)
\ee 
and from the right by $S$
 to obtain the wave function ${\cal Y} (u,
\lambda)$:
\be
{\cal Y} (u, \lambda) = M^{-1} (u) \Psi  (u, \lambda)S = 
 S^{-1}M_0^{-1}( u)  M  (u, \lambda) g^{-1} (\l)S\,.
\lab{def-y}
\ee
${\cal Y} (u, \lambda)$ satisfies, as in \ct{Dubrovin:2001} :
\br
{\cal Y}(u,-\lambda)^T\eta{\cal Y} (u, \lambda)&=&\eta\\
\frac{\pa {\cal Y} (u, \lambda)}{\pa u_i } &=& \l {\cal E}_i (u) {\cal Y} (u, \lambda)
\lab{clayui}\\
\frac{\pa {\cal Y} (u, \lambda)}{\pa x^{\a} } &=& \l C_{\a} (u) {\cal Y} (u, \lambda)
\lab{clayxa}\\
\frac{\pa {\cal Y} (u, \lambda)}{\pa \l} &=& \( {\cal U} (u) + \l^{-1}  \mu
\) {\cal Y} (u, \lambda)
\lab{clayla}
\er
where 
\br
{\cal E}_i (u)  &=& M^{-1}(u) E_{ii} M(u)\lab{defe}\\
{\cal U} (u)  &=& M^{-1}(u) U M(u) \lab{defu}\\
\mu &=& M^{-1}(u) V(u) M(u)=S^{-1}{\cal V}S\lab{defv2}
\er
and 
\be
C_{\a} (u) = \sum_{\b,\g=1}^N c^{\g}_{\a \b} E_{\b \g}
=\sum_{\b,\g,\d=1}^N c_{\a \b \d} \eta^{\d\g}E_{\b \g}\,. 
\lab{defc}
\ee
  
For general $V(u)$, which may contain a nilpotent part, we have $\mu^T\eta+\eta\mu=0$.
Define again ${\cal Y} (u, \lambda)$ by \rf{def-y} then also
\be
{\cal M} (u, \lambda) = M^{-1}(u)  M (u, \lambda) S= I + O(\lambda)
\lab{defm}
\ee
such that ${\cal Y} (u, \lambda)= {\cal M} (u, \lambda) S^{-1}g^{-1} (\lambda)S$ and 
\be 
{\cal M} (u, -\lambda)^T\eta {\cal M} (u, \lambda)=\eta.
\ee
{}From equation \rf{clayla}
one obtains :
\be
g^{-1} \frac{\partial g }{\partial \lambda}
= S{\cal M}^{-1} ( \frac{\partial {\cal M}}{\partial \lambda}
-  {\cal U} {\cal M} -  \frac{\mu}{ \lambda}{\cal M})S^{-1}
\lab{onga}
\ee
Assume, that $g = g_{-} g_{+}$ then
\be
g_{-}^{-1} \frac{\partial g_{-}}{\partial \lambda}  
+ \frac{\partial g_{+} }{\partial \lambda} g_{+}^{-1}
=  g_{+} S{\cal M}^{-1} ( \frac{\partial {\cal M}}{\partial \lambda}
-  {\cal U} {\cal M} -  \frac{\mu}{ \lambda}{\cal M})S^{-1}
g_{+}^{-1}
\lab{ongb}
\ee
Define $G = g_{+} S{\cal M}^{-1}S^{-1}$ then
\be
g_{-}^{-1} \frac{\partial g_{-} }{\partial \lambda}  =
- \frac{1}{\lambda} G_0 S{\mu}S^{-1} G_0^{-1}=
- \frac{1}{\lambda} G_0 {\cal V} G_0^{-1}
\lab{ongmin}
\ee
with matrix $G_0= (g_{+} S{\cal M}^{-1}S^{-1})_0 = (g_{+} )_0$
being orthogonal for a fixed point $g$ 
of the automorphism $\sigma$.

\vskip 10pt \noindent
{\bf Acknowledgements} \\
H.A. is partially supported by NSF (PHY-9820663).

\end{document}